\documentclass[12pt]{article}

\usepackage{amssymb}
\usepackage{graphicx}

\usepackage[hang,small,bf]{caption}
\newcommand{\be}[3]{\begin{equation}  \label{#1#2#3}}
\newcommand{\ee}{\end{equation}}
\newcommand{\ba}{\begin{array}}
\newcommand{\ea}{\end{array}}
\newcommand{\bea}[3]{\begin{eqnarray}  \label{#1#2#3}}
\newcommand{\eea}{\end{eqnarray}}

\newcommand{\ip}{\raise1pt\hbox{\large$\lrcorner$}\,}

\let\Large=\large
\let\large=\normalsize

\def\1{\mathbf 1}

\def\N{{\cal N}}
\newcommand{\haken}{\mathbin{\hbox to 8pt{%
                 \vrule height0.4pt width7pt depth0pt \kern-.4pt
                 \vrule height4pt width0.4pt depth0pt\hss}}}

\renewcommand{\theequation}{\thesection.\arabic{equation}}

\newcommand{\resetcounter}{\setcounter{equation}{0}}  

\begin{document}

\baselineskip=19pt
\parskip=6pt


\thispagestyle{empty}

\begin{flushright}
\hfill{UPR-1111-T}\\
\hfill{LMU-ASC 09/05}\\
\hfill{hep-th/0502154}\\
\hfill{February 17, 2005}

\end{flushright}

\vspace{10pt}

\begin{center}{ {\Large\bf General  Type IIB Fluxes with SU(3) Structures }}

\vspace{35pt}

{ Klaus Behrndt\,$^{1}$ , \ Mirjam Cveti\v c\,$^2$ \ and \ Peng Gao\,$^2$}

\vspace{15pt}

$^1$ 
{\it Arnold-Sommerfeld-Center for Theoretical Physics \\
Department f\"ur Physik, Ludwig-Maximilians-Universit\"at 
M\"unchen,\\
Theresienstra{\ss}e 37, 80333 M\"unchen, Germany
}\\[1mm]

\vspace{8pt}

$^2$ {\it Department of Physics and Astronomy, \\
 University of Pennsylvania, Philadelphia, PA 19104-6396, USA}\\[1mm]

\vspace{40pt}

{ABSTRACT}

\end{center}

\noindent
A supersymmetric vacuum has to obey a set of constraints on fluxes as
well as first order differential equations defined by the
$G$-structures of the internal manifold. We solve these equations for
type IIB supergravity with SU(3) structures. The 6-dimensional
internal manifold has to be complex, the axion/dilaton is in general
non-holomorphic and a cosmological constant is only possible if the
SU(3) structures are broken to SU(2) structures. The general solution
is expressed in terms of one function which is holomorphic in the
three complex coordinates and  if this holomorphic function is constant,
we obtain a flow-type solution and near poles and zeros we find the
so-called type-A and type-B vacuum.

\vfill

\newpage

\tableofcontents

\newpage


\section{Introduction}


A string theory vacuum has to meet a number of requirements to be
consistent with 4-dimensional phenomenology. Not only the moduli,
which are abundant in string theory compactifications, have to be
fixed, but also chiral fermions charged under the gauge group of the
standard model have to present.  Most difficult however, is to achieve
this in a space time with a small positive cosmological constant. A
good starting point are however, the supersymmetric ground states with
a negative or zero cosmological constant. A crucial ingredients for
lifting the moduli space is the presence of background fluxes, which
generate a potential for the scalar fields and there are basically two
approaches to these flux compactifications. If one neglects the back
reaction of the fluxes on the geometry, one can simply perform a
KK-reduction on a Calabi-Yau space yielding the explicit form of the
potential. This is especially reliable as long as one has only RR or
only NS fluxes, see \cite{PS, 053, 240}. The other approach
discusses background fluxes as specific torsion components of the
internal geometry and classifies the vacua with respect to the
corresponding $G$-structures
\cite{Gauntlettetal, Louisetal, LustetalII,056, 140, 650, 058,120}. 
This approach takes the back reaction of the fluxes on the geometry
into account and gives not only the deformed geometry but identifies
also the embedding of the fluxes. It does however not give an
expression for the potential and a discussion of deSitter vacua is not
possible a priori.

To consider type IIB string theory is especially interesting due to
its role for providing supergravity duals of RG-flows and for
providing the basis on most discussions on string cosmology
\cite{840}. {From} the technical viewpoint, fluxes on the IIB side do
not generate the severe back-reaction onto the geometry as it is known
on the type IIA side \cite{780, 175, 888, 042, 041} (see
\cite{055} for type IIA fluxes that also preserve the Calabi-Yau
property). In fact, as we will verify, the internal geometry is always
given by a complex space and it is known that 3-form fluxes can be
turned on while still preserving the Calabi-Yau space \cite{150, GKP,
  145, KachruetalI}.  On the other hand, type IIB supergravity yields
generically a no-scale model related to at least one un-fixed modulus
(typically the volume modulus) and the supersymmetric vacuum is a flat
4-dimensional Minkowski space. The no-scale structure can be lifted by
non-perturbative effects due to brane-instantons that typically
generate a (negative) cosmological constant. Since branes are sources
for fluxes it would be interesting to see whether AdS vacua can also
be obtained from flux compactifications and if so, they are promising
candidates for vacua without (closed string) moduli. Having fixed the
moduli, one can address the other important questions: Are there
chiral fermions and a proper gauge group supported on these vacua? On
the type IIA side, a promising vacuum was presented in \cite{780, 041}
and one might ask whether this vacuum has a mirror dual on the IIB
side. It is also interesting to ask: If we consider general fluxes, is
there still a landscape of string vacua as discussed by Douglas
\cite{051} in traditional CY compactifications? The fluxes on the type
IIA side single out a rather unique vacuum, can this also happen on
the type IIB side?

In the discussion of fluxes, the Ansatz for the Killing spinor plays
an important role and there is a direct link between specific
solutions (as e.g.\ supergravity brane solutions) and the assumptions
for the spinor.  The type-A(ndy) vacuum, for example, is obtained if
the Killing spinor is Majorana-Weyl and describes the NS-type vacuum
of \cite{Strominger} and the prototype supergravity solutions are
NS5-branes.  The S-dual configuration is called type-C vacuum
\cite{155}. For the so-called type-B(ecker) vacuum, one assumes the
Killing spinor to be a direct product of a chiral internal and
external spinor. This was originally introduced in \cite{057} and
adopted in the analysis of Grana and Polchinski \cite{150,145}. The
prototype supergravity solutions in this class are 3- and
7-branes. The most general Ansatz is of course an interpolation, which
we will consider later-on. We should also mention the
type-D(all'Agata) vacuum related to SU(2) structures \cite{135}.

The flow type solutions have been discussed most intensively over the
years and some aspects can be found, for example in
\cite{175,042, 059, 155, 044, 125}.  Interestingly, Frey
\cite{125} was able to solve the equations explicitly in this
class, whereas other explicit solutions are restricted to a sub-class
of fields or torsion classes. In this paper, we could solve all
equations, imposed by supersymmetry and SU(3) structures, without
making any assumptions. The solution preserves four supercharges, ie.\
yields ${{\cal N}=1}$ supersymmetry upon compactification to four
dimensions, the internal space is complex and a cosmological constant
cannot be generated as long as we keep SU(3) structures. The fields
are expressed in terms of one holomorphic function $f=f(z)$, which can
be chosen freely. If it is constant we recover the flow type solutions
whereas zeros and poles of this function can be related to the type-B
and type-A vacua. We solved only the supersymmetry equations and the
solution contains, besides the holomorphic function, one un-fixed
field which, in the spirit of \cite{044}, is the ``master function''
and has to be fixed by the equations of motion or Bianchi
identities. This is nothing specific to our solution, but a well-known
feature of BPS vacua.

The paper is organized as follows. In the next Section, we setup our
conventions and discuss especially the spinor Ansatz and the Killing
spinor equations. In Section 3 we derive the constraints on the fluxes
and differential equations from the supersymmetry variations. The main
part is Section 4, where we first solve the differential equations
explicitly, followed by a discussion of cases related to special
assumptions for the holomorphic function $f(z)$.  We discuss three
cases in more detail: $(i)$ RG-flow for $f=const.$, $(ii)$ special
Hermitian and K\"ahler manifolds and $(iii)$ we expand the solution
around zeros and poles of $f(z)$. In the last section we also comment
on the cosmological constant. Recall, SU(3) structures forbid a
cosmological constant, but as we will show, for SU(2) structures a
cosmological constant can be generated.  In allowing for SU(2)
structures, one has however to keep in mind, that there may appear an
issue in obtaining chiral fermions [similar to the situation on
the type IIA side \cite{100}].


\section{Warp Compactification and Fluxes}

\resetcounter


\subsection{Bosonic Fields}


We use the standard convention of type IIB supergravity \cite{048}
(see also \cite{115, 150}) and combine the bosonic fields into a
complex 1-form $P$, a complex 3-form $G$ and the 5-form $F$
\be008 \ba{rcl} 
P_{M} &=&  {\partial_M B \over 1 -|B|^2} \ , \quad 
B = \frac{1+i\tau}{1-i\tau} \ , \\
G_{(3)} &=& {F_{(3)}-BF_{(3)}^\star \over \sqrt{1-|B|^2}} \ ,\quad
F_{(3)} = d (B_{(2)}+i\, C_{(2)}) \ , \\
F_{(5)}&=& dA_{(4)}-{1\over 8}\,{\rm Im}[(B_{(2)}+i\, C_{(2)})\wedge
F_{(3)}^\star]
\ea \ee
where $\tau = \tau_1 + i \, \tau_2 = C + i \, e^{-\phi}$ combines the axion
$C$ and the dilaton $\phi$. The equations of motions read
\bea265 
{\cal D}^MP_M&=&{1\over24}G_{MNP}G^{MNP}\
, \\
{\cal D}^MG_{MNP}&=&P^M\bar G_{MNP}-{2i\over3}
F_{MNPQR}G^{MQR}\ ,
\\
R_{MN}&=&P_M\bar P_N+\bar P_M
P_N+{1\over6}F_{MLPQR}F_N{}^{LPQR}\nonumber 
\\
& & +{1\over8}(G_M{}^{LP}\bar
G_{NLP}+G_N{}^{LP}\bar G_{MLP}-{1\over6}g_{MN}G_{LPQ}\bar
G^{LPQ}) 
\eea
where the covariant derivative ${\cal D}_M=D_M-iq\,Q_M$ with $D_M$ as the
usual covariant derivative to the Levi-Civita connection and $q$ the
local $U(1)$ charge ($G_{MNP}$ has charge $q=1$, and $P_M$ has the charge
$q=2$) with respect to the $U(1)$ connection
\be007 
 \quad Q_{M}= {{\rm Im}(B \partial_M B^\star) \over 1 -|B|^2}  \ .
\ee
By employing differential forms, the equations of motion can also be
written in a more compact way as
\be298 \ba{rcl} 
d\, {^\star G}&=&iQ\wedge \,{^ \star G} + P\wedge {^\star \bar G}
- 4i \, G\wedge {^\star F} \ , \\ d \, {^\star P} &=& 2iQ\wedge
{^ \star P}+{1\over4} G\wedge {^\star G}  
\ea \ee
and the Bianchi identities for the forms are given by
\be297 \ba{l} 
dQ=-iP\wedge\bar P \quad ,\quad
dP=2iQ\wedge P \ , \
\\
dF={5\over12}i \, G\wedge\bar G \quad ,\quad
dG=iQ\wedge G-P\wedge \bar G \ . 
\ea \ee
Defining a phase $e^{2i\theta}={1+i\bar \tau\over 1-i\tau}$, one finds
moreover
\be255 \ba{rcl} 
Q_M= \partial_M\theta -{\partial_M\tau_1
  \over2\tau_2 } \quad ,\quad P_M=ie^{2i\theta}{\partial_M\tau
  \over2\tau_2} \quad ,\quad
G_{(3)}=i{e^{i\theta}\over\sqrt{\tau_2}}(d A_{(2)}-\tau dB_{(2)})\ .
\ea \ee
The phase $\theta$ drops out from the equations of motion as well as
from the Bianchi identities and can be set to zero yielding the string
theory convention, but note, we are working in the Einstein frame and
thus $G_3$ has the pre-factor $\tau_2^{-1/2} = e^{\phi/2}$. There is
a local U(1) symmetry related to this phase, which we will
discuss below at eq.\ (\ref{134}).

We are interested in vacua, which preserve four supercharges with a
maximal symmetric external space, ie.\ either flat or anti de Sitter
space. The 10-dimensional geometry is therefore a warped product of
the 4-dimensional space-time $X_{1,3}$ and the 6-dimensional internal
space $Y_6$ and we write the metric as
\be001 
ds^2 = e^{-2V(y)} \, g_{\mu\nu} \, dx^{\mu} dx^{\nu} +
e^{2U(y)} \, h_{mn}(y) \, dy^m dy^n 
\ee
where $(g_{\mu\nu},h_{mn})$ are the metrics on $(X_{1,3},Y_6)$ and the warp
factors depend only on the internal coordinates. To preserve 4-d
Poincar$\grave{e}$ invariance, the 3-form field strength $G_{(3)}$ and the
one-form $P$ have non-zero components only inside $Y_6$. On the other hand,
the self-dual 5-form $F_{(5)}$ has to have also components along the
external space, but they are proportional to the four dimensional
Poincar$\grave{e}$ invariant volume form, ie.\footnote{Throughout this
paper, we use $\epsilon_{\mu_1 \mu_2 ... \mu_p}$ for the antisymmetric
tensor and $\varepsilon_{\mu_1 \mu_2 ... \mu_p}$ for the corresponding
tensor density. In other words, $\epsilon_{\mu\nu\lambda\rho}=
e^{-4V}\sqrt{-g}\varepsilon_{\mu\nu\lambda\rho}$ and
$\epsilon_{mnpqrs}=e^{6U}\sqrt{h}\varepsilon_{mnpqrs}$.}
\be002 
\ba{rcl} 
 F_{\mu\nu\lambda\rho m} &=&
 5\, e^{4V}\epsilon_{[\mu\nu\lambda\rho}\,\partial_{m]}Z(y)\  ,\\[1mm]
 F_{mnpqr} &=& -
 e^{4V} \epsilon_{mnpqrs}\,\partial^{s}Z(y)\ . 
\ea \ee
%


\subsection{The Supersymmetry Variations}


The fermionic fields of Type IIB supergravity consists of a
complex Weyl gravitino $\Psi_{M}$ ($\Gamma^{11}\Psi_{M}=-\Psi_{M}$)
and a complex Weyl dilatino $\lambda$ ($\Gamma^{11}\lambda=\lambda$).
A bosonic vacuum, with unbroken supersymmetry requires that the
supersymmetry variations of the fermionic fields vanish yielding
\be006 \ba{rcl} 
0= \delta\lambda &=& \frac{i}{\kappa} \Gamma^{M}P_{M}
 \epsilon^\star- \frac{i}{24}\Gamma^{MNP}G_{MNP} \, \epsilon\ ,
 \label{dilatino}
\\
0= \delta\Psi_{M} &=& \frac{1}{\kappa}\Big(D_{M} -{i\over2} Q_M\Big)
 \epsilon + \frac{i}{480} \Gamma^{M_{1}...M_{5}}
 F_{M_{1}...M_{5}}\Gamma_{M}\, 
\epsilon \\
& & +\frac{1}{96}\Big(\Gamma_{M}\!^{PQR}G_{PQR}-
9\Gamma^{PQ}G_{MPQ} \Big) \epsilon^{\star} \ . 
\label{gravitino} 
\ea \ee
We combined here both Majorana-Weyl supercharges of the same chirality
into one complex Weyl spinor $\epsilon$. Any non-trivial solution of
these eqs.\ corresponds to an unbroken supercharge and $\epsilon$ is
called a Killing spinor. These equations are invariant under the local
U(1) gauge transformation
\be134
\epsilon \rightarrow e^{i g} \epsilon
\quad , \qquad 
\theta \rightarrow \theta + g
\ee
where the phase $\theta$ was introduced at equation (\ref{255}) and we
denoted the corresponding U(1) charge by $q$. This local symmetry is due
to the coset $SL(2,R)/U(1)$ which is parameterized by the scalar fields
of type IIB supergravity and it implies that the phase $\theta$ can be
chosen freely, one can take $\theta =0$ (string theory convention) or
$e^{2i\theta}={1+i\bar \tau\over 1-i\tau}$ (supergravity convention) or
any other value.

As we have done it for the metric and the fluxes, we have also split
the spinors into external and internal components, ie.\ we have to
distinguish between internal and external spinors. Globally
well-defined supersymmetry requires that Killing spinors are singlets
under the structure group of the internal manifold and depending on
the group this provides a classification of supersymmetric vacua.  If
the structure group is $SU(3)$, there can exist only one internal
spinor and for $SU(2)$ we can have two singlet spinors.  The maximal
number would be four corresponding to a trivial structure group.  In
the following, we are only interested in SU(3) structures and hence we
consider for the 10-dimensional spinor the general Ansatz
\be004 \ba{rcl} 
\epsilon &=& a \, \zeta_1 \otimes \chi + b^\star \, 
\,\zeta_2 \otimes \chi^\star \ 
\ea \ee
where $\chi$ is the internal Weyl spinor and $\zeta_{\{1/2\}}$ are two
4-dimensional spinors of opposite chirality (so that $\epsilon$ is
chiral).  We are primarily be interested in vacua preserving only four
supercharges yielding $\N=1$, $D=4$ supersymmetry upon compactification
and it is worth pointing out that the internal $SU(3)$-structure alone
does not guarantee that the vacuum has only ${\cal N}=1$ in four
dimensions.  In fact, one generically gets ${\cal N}=2$ in four
dimensions with the two spinors $\zeta_{\{1/2\}}$. Only if these two
spinors are not independent and obey a projector, the supersymmetry is
reduced to $\N=1$ or four supercharges.  Due to the chirality of these
spinors, we can only impose
\be811
\zeta_2=\zeta_1^\star = \zeta \ .
\ee
Of course, any non-trivial factor appearing in this relation can be
absorbed into the two complex functions $a$ and $b$ which are
introduced for the most general $SU(3)$ invariant Killing
spinor. There are some special cases for the spinor Ansatz, which
played an important role in the literature and which we mentioned
already shortly in the introduction. The type-A vacuum is given by $a
= b^\star$ so that $\epsilon$ becomes a Majorana-Weyl spinor.  This
Ansatz yields especially NS-type vacua and the S-dual configuration is
described by the type-C vacuum. On the other hand, the type-B vacuum
corresponds to $b=0$ (or $a=0$) and the importance of this Ansatz is
due to the fact, that it yields always an internal space that is
K\"ahler, or for constant axion/dilaton Calabi-Yau\footnote{We
refer to K\"ahler spaces as spaces with U(3) holonomy, in contrast to
Calabi-Yau spaces that have SU(3) holonomy.}.

To be concrete, we will use the following chirality conventions
\be005 \ba{rcl} 
\Gamma^{1\!1}\epsilon = -\epsilon \qquad , \qquad
\gamma^5\zeta=\zeta \qquad , \qquad \gamma^7 \chi=-\chi 
\ea \ee
with $\chi$ normalized to $\chi^\dagger \chi=1$. Actually, without
loss of generality, we can take $\chi= e^{-\beta} \chi_0$ with
$\chi_0$ as constant spinor and the phase $\beta$ will drop out in
most of our calculations (any coordinate dependence can be absorbed in
$a$ and $b$).  We now decompose the $\Gamma$-matrices as usual
\be003 \ba{rcl} 
\hat{\Gamma}^{\hat\mu} &=&
\hat{\gamma}^{\hat\mu}\otimes \mathbf{1} \qquad ,  \qquad
\hat\Gamma^{\hat{m}} = \gamma^5 \otimes
\hat\gamma^{\hat{m}} \qquad , \qquad \Gamma^{1\!1} = 
 \gamma^5 \otimes \gamma^7 \ , \\
 \gamma^5 &=& {i\over4!}
 \tilde\epsilon_{\mu\nu\lambda\rho}\gamma^\mu\gamma^\nu
\gamma^\lambda\gamma^\rho
 \qquad , \qquad \gamma^7 = -{i\over6!} \tilde\epsilon_{mnpqrs}
 \gamma^m\gamma^n\gamma^p\gamma^q\gamma^r\gamma^s
\ea \ee
where hatted $\hat\Gamma^{{\hat{M}}}$ and $\hat\gamma^{{\hat{\mu}}}$
have flat tangent space indices and $\tilde{\epsilon}$ denotes tensors
with respect to the un-warped metric. We use the Majorana
representation so that $\hat\Gamma^{\hat\mu}$, $\hat\Gamma^{\hat{m}}$,
$\Gamma^{11}$, $\hat\gamma^{\hat\mu}$ are real and
$\gamma^5$,$\gamma^7$,$\hat\gamma^{\hat{m}}$ are imaginary and
antisymmetric.

The covariant derivative $D_M$ in the gravitino variation refers to the
ten dimensional warped metric $G_{MN}=(e^{-2V}g_{\mu\nu},
e^{2U}h_{mn})$, which is related to the covariant derivatives
$(\nabla_{\mu},\nabla_{m})$ with respect to $(g_{\mu\nu},h_{mn})$ by
\be009 \ba{rcl}
 D_{\mu} &=& (\nabla_{\mu}\otimes\1)-{1\over2}e^{-\!V\!-\!U} 
 (\gamma_{\mu}\gamma^5\otimes\partial V)\  , \\
 D_m &=& \1\otimes(\nabla_m +{1\over2}\gamma_m{}^n\partial_nU)\ .
\ea \ee
Since we do not assume, that the external space is flat, also the
4-dimensional spinor is not necessarily covariantly constant. Instead
we use
\be000 \nabla_\mu \zeta \ = \gamma_\mu \, \bar W \, \zeta^\star 
\ee
and $\bar W = W_1 - i \, W_2$ is related to the 4-dimensional
superpotential or more exactly, its value in the vacuum. The
integrability constraint for this spinor equations requires that the
vacuum is anti deSitter space.  Actually, from 4-dimensional
supergravity we infer that our $W$ is related to the holomorphic
superpotential by ${W \sim e^{K/2} W_{hol}}$, where $K$ is the
K\"ahler potential.

Now, inserting our spinor Ansatz into the variations (\ref{gravitino})
yields
\be014 \ba{rcl}
\delta\lambda &=& ie^{-U} \zeta\otimes(bP\!-\!{a\over24}e^{-2U}G)\chi 
\  \\
& & -ie^{-U}\zeta^\star \otimes (a^\star P\!-\!{b^\star
\over24}e^{-2U}G)\chi^\star \ 
\ea \ee
where $P=\gamma^mP_m$ and $G=\gamma^{mnp}G_{mnp}$.  In evaluating the
supersymmetry variation of the gravitino, we need the field strength
$F_{(5)}$ contracted with $\Gamma$-matrices and find
\be011 \ba{rcl} 
{ i\over 480}\Gamma^{M_{1}...M_{5}}
F_{M_{1}...M_{5}} &=&{1 \over 4}e^{-U} ( \1\otimes\gamma^m\partial_mZ -
 \gamma^5\otimes\gamma^7\gamma^m\partial_mZ)\ \\
&=&{1 \over 2}e^{4V-U} (\1\otimes\partial Z) 
\ea \ee
where $\partial=\gamma^m\partial_m$. Then, the
external supersymmetry variation of the gravitino becomes
\be015 \ba{rcl} 
\delta\Psi_{\mu} &=&  \gamma_\mu\zeta\otimes\Big[W
b^\star\! \chi^\star -{1\over2}e^{-\!V\!-\!U}(a\partial V\!-\!ae^{4V}
\partial Z\!-\!{ b\over48}e^{-2U}G)\chi \Big] + \\
& & \gamma_\mu\zeta^\star\otimes \Big[\bar W  a\chi
+{1\over2}e^{-\!V\!-\!U}(b^\star \partial V\!+\!b^\star e^{4V}\partial
Z\!-\!{a^\star \over48}e^{-2U}G) \chi^\star \Big]\  \ .
\ea \ee
Finally, for the internal gravitino variation we get
\be020 \ba{l} 
\delta\Psi_m = \\
\zeta\otimes
\Big[\nabla_m+{1\over2}(\gamma_m{}^n\partial_nU\!-\!iQ_m\!+\!e^{4V}\partial
Z \gamma_m)+ {b \over 96 a } e^{-2U}(\gamma_m G\!-\!12
G_m)\Big]a\chi + 
\\
\zeta^\star \otimes \Big[\nabla_m+
{1\over2}(\gamma_m{}^n\partial_nU\!-\!iQ_m\!-\!e^{4V}\partial Z
\gamma_m)+ {a^\star \over 96 b^\star} e^{-2U}(\gamma_m
G-12G_m)\Big]b^\star\chi^\star  
\ea \ee
where $G_m=\gamma^{np}G_{mnp}$. Setting these variations to zero gives
us constraints on the background field strengths and in addition
differential equations for the internal Killing spinor, which in turn
require constraints on the geometry of the internal space expressed by
specific non-vanishing torsion components.


\section{Constraints on ${\cal N}=1$ Fluxes and Intrinsic Torsion}

\resetcounter


We will start with the flux constraints, that come from the dilatino
(\ref{014}) as well as from the external gravitino variation
(\ref{015}).


\subsection{Constraints of Fluxes}


Using the above relations and the one that we collected in the
appendix, one gets from $\delta\lambda=0$ (recall, $\zeta$ and
$\zeta^\star$ have opposite chirality)
\be045 \ba{rcl}
\Omega_{mnp}G^{mnp}=\bar \Omega_{mnp}G^{mnp}&=&0 \ , \\
(P_r-iJ_{rs}P^s)+i{a\over8b}e^{-2U}(J_{[mn}h_{p]r}+iJ_{[mn}J_{p]r})
G^{mnp}&=&0  \ , \\
(P_r+iJ_{rs}P^s)-i{b^\star\over8a^\star}e^{-2U}(J_{[mn}h_{p]r}-iJ_{[mn}
 J_{p]r})G^{mnp}&=&0
\ . 
\ea \ee
The constraints in the first line imply that
$G\wedge\Omega=G\wedge\bar\Omega=0$ and hence
$G_{(3,0)}=G_{(0,3)}=0$. By using holomorphic indices these
constraints can be written as
\be047
\ba{rcl}
& &G_{ijk}=G_{\bar i\bar j\bar k}=0 \ , \\
& &P_i={b^\star\over4a^\star}e^{-2U}G_{ij}{}^j\ ,\\
& &P_{\bar i}={a\over4b}e^{-2U} G_{\bar i \bar j}{}^{\bar j}\ .
\ea \ee
With these constraints $\delta\Psi_\mu=0$ yields
\be050 
\ba{rcl}
a\bar W = b^\star W &=&0 \ , \\
{a^\star}^2P_i &=&  2 {b^\star}^2 (\partial_iV+ e^{4V}\partial_iZ)\ ,\\
b^2 P_{\bar i} &=&  2a^2 (\partial_{\bar i}V- e^{4V}\partial_{\bar i}Z)
\ . 
\ea \ee
This has the important consequence that
\be652
W = 0 
\ee
and hence the 4 dimensional cosmological constant is identically zero
and the external space time is always flat Minkowskian.  Note, it {\em
does not} mean that there is no superpotential, it only implies that
the supersymmetric vacuum is given by: $dW = W = 0$.  Below we will
see, that in order to generate a 4-dimensional cosmological constant
one has to consider SU(2) structures.


\subsection{Killing Spinor and Intrinsic Torsion}


Finally, the internal variation $\delta\Psi_m=0$ gives differential
equations for the Killing spinor fixing the torsion components of the
internal manifold.  Using the same relations as before, eq.\ 
(\ref{020}) simplifies to
\be059 \ba{rcl} 
\hat \nabla_m (a\chi) &=&  -{1\over2} \gamma_{mn}(\partial^n U \!+\!
\partial^n V \!-\! 2e^{4V}\partial^n Z)a\chi+{ b\over 8a}
e^{-2U} G_m a\chi \ ,
\\
\hat \nabla_m (b^\star\chi^\star) &=& -{1\over2} \gamma_{mn}(\partial^n U
\!+\! \partial^n V \!+\!2 e^{4V}\partial^n Z)b^\star\chi^\star+{a^\star
\over 8b^\star} e^{-2U} G_m b^\star\chi^\star 
\ea \ee
with $\hat \nabla_m = \nabla_m - {i\over 2} Q_m + {1\over2} \partial_m V$.
These two equations are consistent if
\be062 \ba{rcl}
\partial_m\log({b^\star\over a^\star}) = iQ_m-2i e^{4V} J_m{}^n\partial_n 
 Z +{ia^\star\over8b^\star}e^{-2U}J_{np}G_m{}^{np}
-{ib^\star\over8a^\star}e^{-2U}J_{np}\bar G_m{}^{np} \ .
\ea \ee
And notice that $\nabla_p(\chi^T\!\chi^\star)=0$ requires
\be063 \ba{rcl}
\partial_m\log|a|&=&-{1\over2}\partial_m V-{ib\over16a}e^{-2U} 
J_{np}G_m{}^{np} + {ib^\star\over16a^\star}e^{-2U}J_{np}\bar G_m{}^{np}\  .
\ea \ee
The 3-form flux can be replaced by using the relations (\ref{047}) and
(\ref{050}) and hence we find in complex coordinates
\be277 \ba{rcl} 
\partial_i \log ({b\over a})^\star&=& i Q_{i}+2e^{4V}
\partial_i Z \ ,
\\
\partial_{\bar i} \log ({b\over a})^\star&=& 
iQ_{\bar i}+2(|{b\over a}|^2-
|{a\over b}|^{2})\partial_{\bar i} V+2(|{b\over a}|^2+1+
|{a\over b}|^{2})  e^{4V}\partial_{\bar i}Z \ ,\\
 \partial_i \log|a| + {1\over2}\partial_i V &=& -|{b\over a}|^2
(\partial_{i} V+ e^{4V}\partial_{i}Z)+(\partial_{i} V- 
e^{4V}\partial_{i}Z) \ . 
\ea \ee
Combining the first two equations, we find
\be400 \ba{rcl}
\partial_i \log |{b\over a}|^2 &=& 2(|{b\over a}|^{2}\!+\!
|{a\over b}|^{2}\!+\!2) e^{4V}\partial_i Z+
2(|{b\over a}|^2\!-\!|{a\over b}|^{2})\partial_i V \ , 
\\
\partial_i \log ({a^\star b\over ab^\star}) &=& -2iQ_i+
2 \big(|{b\over a}|^{2}\!+
\!|{a\over b}|^{2}\big) e^{4V}\partial_i 
Z +2\big(|{b\over a}|^2\!-\!|{a\over b}|^{2}\big)\partial_i V \ .
\ea \ee
The first equation can be re-written as
\be405 \ba{l}
\partial_i \log \big(|{b\over a}|^2\!+\!1\big) = 2 
\big(|{b\over a}|^{2}\!+\!1\big) \,
e^{4V}\partial_i Z+2\big( |{b\over a}|^2\!-\!1\big)\, \partial_i V
\ea \ee
and with the third equation in (\ref{277}), we get
\be404 \ba{rcl}
\partial_i \log|a| +{1\over2}\partial_i V=-{1\over2}\partial_i 
\log (|{b\over a}|^2+1)\qquad  \Rightarrow \qquad |a|^2+|b|^2=e^{-V}\ .
\ea \ee
The integration constant for the warp factor $V$ can be absorbed into a
coordinate rescaling in the metric Ansatz. Thus our spinor
becomes  equivalent to \cite{125}
\be992
\epsilon=e^{-{V + i\omega \over 2}} \, 
\Big(\sin\alpha\, [\zeta\otimes\chi] +\cos\alpha\, [
\zeta^\star\otimes\chi^\star] \, \Big)
\ee
or 
\be552
a = e^{-{V +i\omega \over 2}} \sin\alpha \quad , \qquad 
b = e^{-{V -i\omega \over 2}} \cos\alpha \ .
\ee
Note, we absorbed the common phase of $a$ and $b$ into the spinor
($\chi =e^{i\beta} \chi_0$), because this phase drops out in most of
our calculations.  In addition, using the U(1) symmetry (\ref{134}),
one can always set $\omega = 0$ (or fix the phase $\theta$).  We will
explore these equations further in the next section.

Since the internal spinor is covariantly constant only for specific
fluxes, we have to compensate the fluxes by intrinsic torsion components
of the internal manifold. In other words, the fluxes deform the internal
geometry in a way dictated by the non-vanishing torsion components. A
real 6-dimensional space with $SU(3)$ structures $J$ and $\Omega$ is
classified by five torsion classes ${\cal W}_{1,2,3,4,5}$, which are
defined by \cite{340}
\be060 \ba{rcl}
 dJ&=&{3i\over4}({\cal W}_1\bar \Omega-
 \bar {\cal W}_1\Omega)+{\cal W}_3+J\wedge{\cal W}_4\ ,\\
 d\Omega&=&{\cal W}_1J\wedge J+J\wedge {\cal
W}_2+\Omega\wedge{\cal W}_5
\ea \ee
with 
\be061
\ba{rcl}
{\cal W}_1\leftrightarrow (dJ)^{(3,0)}\quad ,\quad {\cal W}_2\leftrightarrow 
(d\Omega)^{(2,2)}_0 
\quad ,\quad {\cal W}_3 \leftrightarrow (dJ)^{(2,1)}_0 \ , \\
{\cal W}_4={1\over2}J\ip dJ\leftrightarrow (dJ)^{(1,0)} \quad ,\quad 
{\cal W}_5={1\over2}({\cal R}e\Omega)\ip d({\cal R}e\Omega)
\leftrightarrow (d\Omega)^{(3,1)} \ .
\ea
\ee
The subscript $0$ denotes the irreducible $SU(3)$ representation with the
trace part removed and the operation $\ip$ is defined as\footnote{Also
notice that $(\mathbf{d}\alpha)_{m_1...m_{p+1}}=
(p+1)\partial_{[m_1}\alpha_{m_2...m_{p+1}]}$.}
\[
(\alpha\ip\beta)={1\over p!}(\alpha\cdot\beta)
\]
[here we take $\alpha$ to be the lower form of rank $p$].
Due to the differential equations (\ref{059}) $J$ and $\Omega$ are not
closed and we find
\be093
\ba{l}
\partial_{[m}J_{np]}=
J_{[mn} M_{p]} +{ib\over4a}e^{-2U} \Big( J_{[m}{}^rJ_n^sG_{p]rs}-
G_{mnp}-2iJ_{[m}{}^lG_{l|np]} \Big) +c.c.
\ , \\
\partial_{[m}\Omega_{npq]}=\Omega_{[mnp} N_{p]} \ ,
\ea \ee
with the two vectors: 
\[
\ba{l}
  M= -d\log|a| -{5 \over 2} dV -2 dU +4 e^{4V} dZ
  -{ib \over 8a} e^{-2U} G \ip J \\
  N= 2 d[\log|a|\!+ 2V\!+\! {i \over 2} (\omega + \beta) ]-iQ +
3 (dU - 2e^{4V}dZ) +{ib \over 4a}   e^{-2U} G \ip J
\ea
\]
And therefore the torsion classes read
\be094 \ba{rcl}
{\cal W}_1 &=& 0\ , \\
{\cal W}_2 &=& 0\ ,\\
({\cal W}_3)_{mnp} &=&
{ib\over4a}e^{-2U}\Big(G_{mnp}\!-\!J_{[m}{}^r
J_n^sG_{p]rs}\!-\!{1\over2}J_{rs}J_{[mn}G_{p]}{}^{rs}\Big)
+c.c. \
 ,\\
({\cal W}_4)_p &=& M_p-{b\over 8a}
e^{-2U}J_p{}^lG_{lmn}J^{mn}+ c.c. \
\ , \\
({\cal W}_5)_p &=& 
 {1\over2}\Big[\delta_p{}^l\!-\!iJ_p{}^l\Big]
N_l + cc \ .
\ea \ee
Because ${\cal W}_1={\cal W}_2=0$, the internal space has to be a
complex manifold, i.e. the almost complex structure $J_m{}^n$ is
integrable. This result has also been found in \cite{175}, but since
they assumed a flat Minkowski vacuum, it was not clear if a non-zero
cosmological constant could have been compensated by ${\cal W}_1$ for
example, which is exactly the case on the type IIA side \cite{780,
041}. For generic background fluxes, the intrinsic torsion lies
therefore in $ ({\cal W}_3\oplus{\cal W}_4 \oplus{\cal W}_5)$, with
${\cal W}_3$ purely supported by the $(2,1)$
primitive\footnote{Primitive means $J\ip G_{(3)}=0$.} components of
$G_{(3)}$ and ${\cal W}_{\{4/5\}}$ are related to a non-trivial
axion/dilaton and warping.  One can easily check that our expressions
reduces to the well known results in the corresponding special
limits. For example, when $ab=0$, ${\cal W}_3$ and ${\cal W}_4$ vanish
if we let $U=V$, and ${\cal W}_5=-i(Q +d\beta) $ characterizes a
K\"ahler space. If $b=\pm a$, on the other hand, $2{\cal W}_4+{\cal
W}_5=0$ if we let $U+V=0$. Let us also remark, that the combination
$3{\cal W}_4+2{\cal W}_5$ is independent of $U$ as it should be,
because $U$ is only a conformal rescaling.

In complex coordinates the expressions simplify to
\be299 \ba{l}
 {\cal W}_3{}_{ij\bar k }={ib\over4a}e^{-2U} 
(G_{ij\bar k}-{3\over2}h_{[i\bar k}G_{j]l}{}^l) \ ,\\
{\cal W}_4{}_i=\partial_i\log(\cos\alpha)-2 ( \partial_iU +  \partial_iV
- 2 e^{4V}\partial_iZ) \ ,  \\
{\cal W}_5{}_i=-i\, ( Q_i +  \partial_i\omega + \partial_i \beta) + \partial_i 
\log\cos^2\alpha + 3 \, \partial_iU + 7 \, \partial_iV-
10 \, e^{4V}\partial_iZ  
\ea \ee
where we used eqs.\ (\ref{047}), (\ref{050}), (\ref{405}) and (\ref{992}).

Observe that all constraints required by supersymmetry do not involve
the internal warp factor $U$ and we may thus simplify our torsion
classes by imposing relations between $U$ and the other
functions. Remember however a change in $U$ does change the solution
by a conformal rescaling of the internal metric. In what follows, by
internal geometry we always mean the non-warped metric, so a
Calabi-Yau space really means the internal space equipped with the
warped metric is conformal to a Calabi-Yau and so on.


\section{Solutions of SUSY Constraints}

\resetcounter


To simplify the calculations we use the symmetry (\ref{134}) to set in
the following
\[ 
\omega = 0   
\]
(equivalently, one could also fix the phase $\theta$). We will start by
solving the supersymmetry constraint equations followed by a discussion
of special cases/limits.


\subsection{Solving the supersymmetry equations}


Using the parameterization from (\ref{552}) we find from the first
equation in (\ref{400})
\[
4 \, \partial_i Z = \partial_i \Big( e^{-4V} \cos 2\alpha \Big) 
\]
and hence
\be911
Z -Z_0 =  {1 \over 4} \, e^{-4V} \cos 2\alpha\ .
\ee
To keep the notation simple, we set in the following: $Z_0 = 0$.
The second equation in (\ref{400}) can be written as
\be625
\ba{rcl}
i \, Q_i  &=& 2 \cos 2\alpha \; \partial_i V
+ \Big( {2 \over \sin 2\alpha} - \sin 2\alpha \Big) \, \partial_i \alpha \\
&=& - \cos 2\alpha \, \partial_i \log \sqrt{Z} + \partial_i \log \cot\alpha \ .
\ea
\ee
Next, the equations in (\ref{050}) become
\[
\ba{rcl}
P_i &=& 4 \cos^2\alpha \, \partial_i V + 2 {\cos^3\alpha \over \sin\alpha}
   \, \partial_i \alpha \ , \\
P_{\bar i} &=& 4 \sin^2\alpha \, \partial_{\bar i} V - 
2 {\sin^3\alpha \over \cos\alpha}   \, \partial_{\bar i} \alpha 
\ea
\]
and therefore
\be729
\ba{rcl}
(P + P^\star)_i &=& - \partial_i \log |\tilde Z|  
 \ ,\\
(P-P^\star)_i &=& - \cos 2\alpha \, \partial_i \log |\tilde Z| + 
\partial_i \cos 2\alpha \, = \, 2i \, Q_i
\ea \ee
where we used the expression for $Q_i$ as found in (\ref{625}) and
introduced
\be611
\textstyle{
\tilde Z = Z \; {\sin^2 2\alpha \over \cos 2 \alpha} \ .
}
\ee
The vector $P$ can now be written as
\be242
\textstyle{
P = -{1 \over 2} d\,  \log |\tilde Z| \, 
+ i \, Q 
}
\ee
and from the Bianchi identities in (\ref{297}) we get
\be882
\ba{l}
dQ \equiv - i P \wedge \bar P = -  Q \wedge d \log |\tilde Z|
\ea
\ee
which can formally be solved by: $Q = \tilde Z^{-1} \, Q_0$ where $Q_0$
can be any closed 1-form.  With this expression for $dQ$, one can also
verify that the Bianchi identity for $P$, $dP = 2i Q \wedge P$, is
identical fulfilled for $P$ as given in (\ref{242}).

On the other hand, using the expression for the vector $P$ from eq.\
(\ref{255}), we find
\[
P = {1 \over 2} \Big( - e^\phi \, \sin 2\theta \,  d\tau_1 +
\cos 2\theta \, d\phi \Big) + {i \over 2}  \, \Big( e^\phi\, \cos 2\theta 
\,  d\tau_1 + \sin 2\theta \, d\phi \Big) \ .
\]
In comparing these two expressions for $P$ we get two equations.  Taking
$Q$ from eq.\ (\ref{255}), we find from equating the imaginary parts
\[
0=e^\phi \cos\theta \, \partial_i \tau_1 + \sin\theta \, \partial_i \phi -
{\partial_i \theta \over \cos\theta} 
\]
whose integral is
\be220
\tau_1 = c_0 + {\sin \theta \over \cos\theta  } \, e^{-\phi} \ .
\ee
Then, from equating the real parts we find
\be822
e^{-(\phi-\phi_0)} =  \tilde Z \, \cos^2\theta   \ .
\ee
We have now everything expressed in terms of $\theta$, $\alpha$ and $Z$ and
these functions have to solve the eqs.\
\[
2i \, Q_i \equiv 2i\,  
\Big(\, \partial_i \theta - {\partial_i \tau_1 \over 2 \tau_2}\, \Big)
=- \cos 2\alpha \, \partial_i \log |\tilde Z| + 
\partial_i \cos 2\alpha \ .
\]
Inserting our result for $\tau$ we find
\[
\ba{l}
2i\, Q_i = i \, \partial_i \tan\theta - i \, \tan\theta \, 
\partial_i \log |\tilde Z|  
\ea
\]
and thus
\be999
\partial_i \Big(\, {\tilde Z \over \cos 2\alpha - i \tan\theta } 
\, \Big) = \partial_{i} f^\star = 0 \ .
\ee
By complex conjugation, we infer that the complex function
\[
f={\tilde Z \over \cos 2\alpha + i \tan\theta}
\]
has to be holomorphic in the three complex coordinates parameterizing
the internal space.  Separating the complex function $f$ into real and
imaginary part, this equation can then also be written as
\be543
\ba{l}
0={\rm Re} f \, \tan\theta + {\rm Im} f \, \cos 2\alpha \ , \\
\tilde Z =  {|f|^2 \over {\rm Im} f} \, \tan\theta
= - {|f|^2 \over {\rm Re}f} \, \cos 2\alpha 
\ea
\ee
and the final solution reads 
\be997
\ba{rcl}
\tau &=& c_0 + i\, e^{-\phi_0}\, { |f|^2 \, \cos 2\alpha \over 
f \, \sin^2\alpha   + f^\star \, \cos^2\alpha} \ , \\ 
e^{4V}&=&  {{\rm Re} f \over 4 |f|^2}\, {\sin^2 2\alpha \over \cos 2\alpha}
 \ ,
 \\
Z &=& {|f|^2 \over {\rm Re} f} \, {\cos^2 2\alpha \over \sin^2 2\alpha} \ .
\ea
\ee
Recall, we dropped the constant $Z_0$, which can be re-introduced by
$Z \rightarrow Z - Z_0$ and similarly, by replacing $\theta
\rightarrow \theta + \omega$ we can again write it in  a $U(1)$-gauge 
invariant form [with respect to local symmetry (\ref{134})].

So, we were able to integrate the first order differential equations
required by supersymmetry and the solution is parameterized by one
holomorphic function $f$. But we have to keep in mind, that in
addition, also the Bianchi identities and the equations of motion for
the form fields have to be satisfied in order to have really a BPS
solution. Note, in total we have to fix five function, the four
bosonic fields on the lhs of (\ref{997}) plus $\alpha$ which fixes the
spinor ($\omega$ can be gauged away). The eqs.\ (\ref{997}) give four
algebraic relations and hence one function remains free (the ``master
function'' as discussed in \cite{044}), which has to be fixed by the
equations of motion -- as it is very typical for supersymmetric
constraints.


\subsection{The $SU(3)$ Flow}


The simplest case is to assume that all functions depend only on one
real coordinate, say $r$, which is the case for a flow-type solution
which have been widely discussed already in the literature \cite{043,
044, 125, 042}. Since there are no holomorphic functions depending
only on one real coordinate we have to set
\[
f = constant
\]
providing two real integration constants. It is straightforward to
verify that in this case our solution as given in (\ref{997}) is
equivalent to the results obtained in \cite{125}. There are two
special cases: $(i)$ ${\rm Im} f =0$ or $(ii)$ ${\rm Re} f = 0$, where
some fields become trivial: for case $(i)$ it is the axion whereas for
$(ii)$ the 5-form flux vanishes, ie.
\[
\ba{l}
{\rm Im} f = 0 \, : \qquad \tau_1 = c_0 	 \ ,    \\
{\rm Re} f = 0 \, : \qquad  Z=0 \ .
\ea
\]
Note, for this choice of $f$ one cannot add, in addition to the
fluxes, 7-branes or 3-branes and in the following we will assume that
$f$ is a generic complex constant.  Since all fields depend only on
one real coordinate, we find trivially
\[
dQ = -i P \wedge \bar P = 0 \qquad , \qquad Q \wedge P = 0
\]
and using the $U(1)$ symmetry (\ref{134}) we can also set $Q\equiv 0$.  In
addition, this case requires $dZ \wedge dV = 0$ and hence we can
always impose ${\cal W}_4 = 0$ by choosing an appropriate conformal 
factor of the internal metric, ie.\ to fix the function $U$.

Recall, the (algebraic) relations in (\ref{997}) fix four of the five
independent function and the remaining one has to be fixed by imposing
the Bianchi identities and/or equations of motion. For this it is
suggestive to take the angle $\alpha$, but we can also take the 5-form
flux function $Z$ as the ``master function'', because this function can
be fixed by the 5-form Bianchi identity (the equations for the 3-form are
trivially solved for this case)
\be921
dF \, = \, { 5i \over 12} \, G \wedge \bar G \ .
\ee
The conformal factor of the internal metric can now be chosen in a way
that the lhs of this equation becomes a (curved space) Laplacian on $Z$.
This is the case for $U = V$ and we can now distinguish the two
possibilities whether the rhs is zero or not.

If $G \wedge \bar G =0 $, the function $Z$ has to be harmonic on the 6-d
internal space. If $Z$ is constant also $\alpha$ is constant and the
solution becomes trivial, ie.\ the internal space is Calabi-Yau. If it is
not constant, the real harmonic function $Z$ has to have a singularity.
Since $e^{-4V} \rightarrow \infty$ at this point, the warp factor blows
up and from the AdS/CFT perspective this point corresponds to the UV
regime. On the other hand, if $Z$ vanishes (or $Z = Z_0$ if we
re-introduce the integration constant), also the warp factor goes to zero
and we reach the IR regime.  Depending on the concrete internal space,
these can be regular fixed points corresponding to an $AdS_5 \times X_5$
geometry, with $X_5$ as a 5-dimensional Einstein space (or products
thereof). Thus, the UV-regime corresponds $\alpha
\simeq 0$ and $e^{- 4V} \simeq 4 Z \simeq  {|f|^2 \over {\rm Re} f
\, \alpha^2}$. Going back to the spinor equation (\ref{059}), in this
regime the spinor $\chi$ becomes covariantly constant and the internal
space becomes Calabi-Yau. Since $U=V$, the point $\alpha \simeq 0$ can
only be regular, if the Calabi-Yau has a conical singularity with $e^{2U}
\simeq \alpha \, r^2 = constant.$, where $r$ is the radial coordinate
of the cone. This yields a regular UV fixed point and $X_5$ has to be
an Einstein-Sasaki space as required by supersymmetry; for an
explicit example see \cite{666}. A prototype example is the $AdS_5
\times S^5$ geometry appearing in the near horizon limit of D3-branes
and we reach this geometry in our setup in the limit
\be772  
f \rightarrow \lambda^2 f   \ , \quad
\alpha \rightarrow \lambda \, \alpha 
\ , \quad  e^{\phi_0} \rightarrow
\lambda^2 \, e^{\phi_0}
\qquad {\rm and} \qquad \lambda \rightarrow
0 \ .
\ee
As consequence, only one chirality of the internal spinor contributes
to the 10-d Killing spinor ($a = 0$ and $b =1$) and $\tau = c_0 + i\,
e^{-\phi_0} \, f $ (and $\theta$ becomes the phase of $f$), which is
the type-B vacuum discussed in \cite{150, GKP}.  Note, the 3-form flux
does not need to vanish, the primitive components of (1,2)-type can
still be non-zero fixing the complex structure moduli of the
Calabi-Yau space.  Therefore, in the UV regime, one can always approximate
the internal geometry by a Calabi-Yau space. In the IR regime on the
other hand, the situation is more involved. It is Calabi-Yau only if
we take the limit (\ref{772}). If we do not re-scale $f$, it
corresponds to $\cos 2\alpha \simeq 0$ and hence $\sin^2 \alpha \simeq
\cos^2 \alpha$ ($a = \pm b^\star$) and the spinor does not
become covariantly constant. {From} (\ref{997}) follows, that we are in a
strongly coupled regime ($e^\phi \rightarrow \infty$) and far off a
Calabi-Yau approximation.  Actually, only for the Calabi-Yau case, we
have control over the IR regime\footnote{Although, in the spirit of
F-theory, this may also be an indication of a de-compactification and the
higher-dimensional theory does not need to be singular.}.  See also
section 4.4 for a related discussion.

Let us also comment on the case if: $G \wedge \bar G \neq 0$. This is
only possible if the 3-form $G$ has primitive components, because all
non-primitive components of $G$, given by $G \wedge J$, are along $dr$
and hence do not contribute to $G \wedge \bar G$. In the case of ${\cal
W}_3 \neq 0$ the space cannot be Calabi-Yau. If one can find, for a
given space, an appropriate 3-form which satisfies the Bianchi
identities and the equations of motion, the function $Z$ can have a
regular maximum (ie.\ no poles anymore), which also represents the
endpoint of the flow ($dZ = dV = d\tau =0$) and is reminiscent to the
transgression mechanism discussed in \cite{045}. Since the warp factor
is finite at this point, this flow terminates on a 5-dimensional flat
space time (instead of an $AdS_5$ fixed point discussed before) and
using results from AdS/CFT, we expect a confining gauge theory dual at
the endpoint of this flow (because Wilson loop calculation show the
area law behavior in this case).


\subsection{Special Hermitian or K\"ahler  spaces}


Next, we will allow for a general holomorphic function $f$ and comment
on the examples of special Hermitian manifolds. Special Hermitian
manifolds are complex half-flat manifolds\footnote{These are
6-dimensional manifolds with $SU(3)$-structure and intrinsic torsion
satisfying ${\cal W}_4={\cal W}_5=0$ and $d({\cal I}\!m\Omega)=0$.
They can be lifted to seven dimensional manifolds with $G_2$-holonomy
by Hitchin's construction, where only the special Hermitian manifolds
are also complex.} and type IIB compactified on half-flat manifolds
arises as mirror symmetric partner of type IIA on Calabi-Yau manifolds
with NS-NS $H_{(3)}$ fluxes \cite{Louisetal}. Both compactifications
usually retain ${\cal N}=2$ supersymmetries in four dimensions.

Observe the condition ${\cal W}_4=0$ implies that $dV \wedge dZ=0$
and we can solve the equation for the function $U$.
Then, we find in the gauge $\omega = 0$
\be333
0={\cal W}_5 = i\,Q - {7\over2}\, \partial \log(\sin\alpha) 
- 4\, \partial V +  4 \, e^{4V}\partial Z 
\ee
where we have taken here only the holomorphic component.  Inserting
for $Q$ the expression as given in eq.\ (\ref{625}) and using the
solution for $Z$ in (\ref{911}), we get a differential equation for
$V=V(\alpha)$. The solution $V = V(\alpha)$ has to be compared
with the one given before in eq.\ (\ref{997}) and hence yields a
relation between $\alpha$ and the holomorphic function $f$.
 
If we do not impose eq.\ (\ref{333}), but instead require the
vanishing of ${\cal W}_3$, the internal space is K\"ahler with ${\cal
W}_5$ as the K\"ahler connection. The simplest possibility has been
discussed by Grana and Polchinski \cite{145}, where the 10-dimensional
spinor factorizes and only one chirality of the internal spinor
contributes ($a=0$). This exactly corresponds to the limit that we
discuss at eq.\ (\ref{772}) with the axion/dilaton
\be661
\tau = c_0 + i \, e^{-\phi_0}\, f
\ee
i.e.\ it is a holomorphic function of the internal coordinates
(recall $f = f(z^i)$) and the 3-form can only have primitive components
of (1,2) type. The warp factor and the 5-form flux become in this
limit
\be985
e^{-4V} = 4 Z = { |f|^2 \over {\rm Re} f \, \alpha^2 }
\ee
which ensures that ${\cal W}_4 = 0$ if we take $U=V$ (recall $\sin\alpha
\rightarrow 0$ in this limit) and the internal space becomes conformal to
a K\"ahler space with the K\"ahler connection (in the gauge $\omega = 0$)
\[
{\cal W}_5 = -i \, (Q +d \beta) \ .
\]
Since the K\"ahler form factorizes ($dQ = -i P \wedge \bar P$), these
can only be very specific K\"ahler spaces and one may wonder, whether
more general spaces with non-holomorphic dilaton are possible.  We do
not want to discuss this question here in detail, let us only mention
that, imposing ${\cal W}_4=0$ implies that the rhs of (\ref{333}) can
be written as
\[
{\cal W}_5 = i \, \partial K
\]
with $K$ as the K\"ahler potential and ${\cal W}_5$ the K\"ahler connection
(recall $dV \wedge dZ = 0$ in order to have ${\cal W}_4=0$).


\subsection{Special limits of the general solution}


Most explicitly known solutions have an internal space, where all
possible torsion classes are non-trivial and the mathematical
classification of spaces with respect to non-vanishing torsion classes,
has no obvious physical interpretation.  In physics literature the
solutions are typically classified with respect to the spinor Ansatz,
e.g.\ whether the 10-dimensional Killing spinor is only Weyl (e.g.\
$\sin\alpha = 0$) or if the spinor is Majorana-Weyl (e.g.\ $\sin\alpha =
\cos\alpha$). Different spinor Ans\"atze correspond to different
solutions, as e.g.\ different brane solutions in supergravity.

The general solution, which we found, is parameterized by a holomorphic
function $f$ and we discussed already in detail the case of constant $f$,
which is the only regular case if the internal space is compact. Other
solutions can be classified with respect to the singularities and zeros
of $f$. If we write
\[
f(z^i) = {p(z^i) \over q(z^i)}
\]
where $p$ and $q$ are some polynomials in the complex coordinates
$z^i$ of different degree. To get a regular solution around a zero of
$p$ or $q$, the other fields have to behave appropriately.

Let us first discuss the solution around zeros of $f$.  Denoting a zero
of the polynomial $p(z^i)$ by $z_p$ , a (regular) expansion around
$z=z_p$ can be identified with the limit (\ref{772}) if we
take\footnote{The corresponding phase gives a shift to $\theta$ and can be
gauged away due to the symmetry (\ref{134}).}: $\lambda^2 = |z -z_p|$.
This (vanishing) parameter drops out and the fields are given in eqs.\
(\ref{661}) and (\ref{985}). Recall, the solution becomes the B-type
vacuum and the supergravity solution in this class describes intersecting
7-branes in the background of a 5-form flux and a primitive 3-form flux
of type $(1,2)$. The resulting geometry is a (special) K\"ahler space.
If the fluxes vanish, ie.\ especially for $Z=const.$ yielding $\alpha^2
\sim |f|^2/{\rm Re}f$, the only non-trivial field is the axion/dilaton
and one can follow the discussion in \cite{046,047}, where the
singularities of the holomorphic function $f$ are related to 7-branes,
that generate a deficit angle and if one has 24 branes, the space
becomes smooth and compact. By considering $D_4$ singularities, one
can also setup an orientifold with D7-branes and coincident O7-planes;
see \cite{999}. Note, if there is no 5-form flux, there is also no
warping ($V$ and $U$ become constant), which one should expect for a
7-brane background in the Einstein frame.

In the same way we can also explore the solution around zeros of the
polynomial $q(z^i)$ giving poles for the holomorphic function $f$.
As before, we are asking for a regular expansion around this point, 
but this time we consider
\be131
\textstyle{
f \rightarrow \lambda^{-2} f \ , \quad \alpha \rightarrow 
( \, {\pi \over 4} - \lambda^2 \, ) \, \alpha \ , \quad
\theta \rightarrow \lambda^2 \, \theta
\qquad {\rm and} 
\qquad \lambda \rightarrow 0   
}
\ee
and obtain
\be922
\ba{rcl}
e^{4V} &=&  {{\rm Re} f \over 8 |f|^2}\,  {1 \over \alpha} \ , \\
Z &=& 0 \ , \\
\tau &=&  c_0 + {i \over 8} \, e^{-(\phi_0 + 4V)} \ .
\ea\ee
Since now $a = b^\star$ we get the A-type vacuum, which has been
analyzed in \cite{Strominger}. All RR fields are trivial up to the
primitive 3-form flux of $(1,2)$ type, which one can always turn
on. The dual (2,1)-forms have to vanish because the Bianchi identity
for the 5-form requires $G \wedge \bar G = 0$ and the non-primitive
part is of NS-type, since $G\wedge J$ is real in this limit.  A
prototype supergravity solution would be the near-horizon geometry of
NS5-branes with a constant warp factor in the string frame metric
($ds_{str}^2 = e^{\phi/2} ds^2 \sim e^{2V} ds^2$), but the dilaton is
not stabilized.

In summary, around zeros of the holomorphic function, the solution is
equivalent to the type-B vacuum, whereas close to a pole of $f$ one
reaches the type-A vacuum.


\section{Comments on the 4-d Cosmological Constant}


The value of the superpotential in the vacuum gives a mass term for
the gravitino and hence enters also the spinor equation (\ref{000}).
We have seen that supersymmetry preserving background fluxes cannot
contribute to $W$, as long as the internal manifold has SU(3)
structures. The external 4-d space time will always be Minkowski. In
order to generate a cosmological constant, one has to consider an
internal manifold with SU(2) structures and let us briefly discuss
some aspects here; see also \cite{100, 135}. In this case, the
internal spinors have to be singlets only under SU(2) and hence there
are two independent Weyl spinors $\chi_1$ and $\chi_2$.  Equivalently,
one can define a real unit vector $v$ globally, which relates the two
spinors by
\be501 \ba{rcl} \chi_2=\gamma_v\chi_1=v_m\gamma^m\chi_1 \ . \ea
\ee
Without loss in generality, we set $\gamma^7\chi_1=-\chi_1$, and then
$\gamma^7\chi_2=\chi_2$ and $\chi_1$ can be taken as the SU(3) singlet
spinor that we used before [of course, $\chi_2$ transforms then under
SU(3) and is a singlet only under SU(2)]. With the notation of
$\chi={\chi_1 \choose \chi_2}$, this is
$\gamma^7\chi_i=-(\sigma_3\chi)_i$, where $\sigma_k$ the usual Pauli
matrices. Decomposing the chiral 10-d supersymmetry transformation
parameters, we get two 4-d Weyl spinors $\zeta_1$ and $\zeta_2$, or
$\zeta={\zeta_1 \choose \zeta_2}$, with chiralities
$\gamma^5\zeta_i=(\sigma_3\zeta)_i$.  In order to get ${\cal N}=1$,
$D$=4 supersymmetric backgrounds, we project out half of the spinor
components of the two 4-d spinors by imposing: 
$\zeta_i=(\sigma_1\zeta^\star)_i$ [due to the chiral choice, this is
the only possibility].  Now, the 10-d Killing spinor decomposes as
\be502 \ba{rcl}
\epsilon=f_1\,\zeta_1\!\otimes\!\chi_1+f_2\,\zeta_2\!
\otimes\!\chi_2+f_3\,\zeta_1\!
\otimes\!\chi_2^\star+f_4\,\zeta_2\!\otimes\!\chi_1^\star
\ea \ee
where $f_i$ are globally defined complex functions, which are the
analogs of the function $a$ and $b$ used before and the spinors are
normalized to
$\chi_i^\dagger\chi_j=\zeta_i^\dagger\zeta_j=\delta_{ij}$.  Now the
superpotential is in general a $2\times2$ matrix denoted $\hat W$,
\be503 \ba{rcl} \nabla_{\mu}\zeta_i=\gamma_{\mu}\hat
W_{ij}\zeta^\star{}^j \ea \ee
However, since $\zeta_i$ are chiral and $\zeta_1^\star=\zeta_2$,
we find that $\hat W$ is diagonal and $\hat W_{11}=\hat
W_{22}^\star=\bar W$. Here as before we defined $\bar
W=W^\star=W_1-iW_2$. 

Now, the calculation can be repeated in the same way as we were doing
it before. Since this is very involved, let us only show here that now
$W$ can be non-zero in the vacuum. For this, it is enough to consider
the simple case, which is the analog of the type-B vacuum, ie.\ where
the spinor is a direct product
\[
\epsilon = \zeta \otimes \chi
\]
with $\zeta_1 \equiv \zeta$ and $\chi = f_1 \chi_1 + f_3
\chi_2^\star$ (or $f_2 = f_4 =0$).  Using the above conventions and
taking the spinor product $(\chi^T \, \delta\Psi_\mu)$, we find from
the vanishing of this expression that
\[
\bar W = {1 \over 16} \, e^{-V -3U} \,  (\chi^T G \chi) 
= {1 \over 8} \, e^{-V -3U} \,f_1 f_3\    (\chi_2^\dagger \, G \, \chi_1) \ .
\]
Inserting $\chi_2 = v_m \gamma^m \chi_1$ and taking $\chi_1$ as the
SU(3) singlet spinor, satisfying the relations from the appendix, one
gets an explicit form of $W$. This is straightforward and we do not
want to discuss further details here.

Note the importance of the vector field, which implies that the
internal space has to admit a complex fibration over a 4-d base
space. As long as one does not take into account brane sources, this
vector field has to be globally defined, which severely constraints
the geometry of the internal space. On the other hand, if we allow for
singularities, eg.\ related to 7-branes wrapping the 4-d base, also
the spinors and the vector field do not need to be globally defined
and to make the supersymmetry nevertheless well-defined one has to add
explicitly the brane sources. This is consistent with the statements,
how a cosmological constant can be generated by wrapped branes
\cite{999}.


\appendix

\section*{Appendix}

\renewcommand{\theequation}{A.\arabic{equation}}


On a real 6-dimensional space, an $SU(3)$ structure is defined by a
2-form $J$ and a 3-form $\Omega$ which are globally well-defined and
are in one-to-one correspondence to an $SU(3)$ singlet spinor $\eta$.
If this spinor (and therewith the forms) is covariantly constant, the
space has $SU(3)$ holonomy, which is however in general not the case
and the deviation is measured by non-zero torsion components. 
As consequence of the $SU(3)$ singlet property, this
spinor satisfies
\be030 \ba{rcl} (\gamma_m-iJ_{mn}\gamma^n)\chi &=& 0 \ ,  \
\\
(\gamma_{mn}+iJ_{mn})\chi &=&
{i\over2} e^{2i\beta}\, \Omega_{mnp}\gamma^p\chi^\star \ ,
\\
(\gamma_{mnp}+3iJ_{[mn}\gamma_{p]})\chi &=&
i e^{2i\beta}\, \Omega_{mnp}\chi^\star \ . \ea \ee 
The first relation is a set of three projector equations for the
$SU(3)$-invariant spinor. Only two of them are independent and,
together with $\zeta_2=\zeta_1^\star$ in (\ref{811}), reduce the
number of supersymmetries to ${\cal N}=1$ in four dimensions. The
fundamental 2-form $J$ and the 3-form $\Omega$ are defined as spinor
bilinears by
\be040 \ba{rcl} 
\chi^T\gamma_{mn}\chi^\star =iJ_{mn} \quad , \quad
\chi^T\gamma_{mnp}\chi=i\, e^{2i\beta} \, \Omega_{mnp}  \ .
\ea \ee
Both are $SU(3)$ singlets and these are the only non-vanishing forms
constructed from $\chi$, because any other form, especially a vector,
would transform under $SU(3) \subset SO(6)$. They satisfy in addition the
following relations (which follow directly from Fierz identities)
\[
\ba{l}
J\wedge \Omega=0 \ , \\
\Omega\wedge \bar \Omega=-{4i\over 3} J\wedge J\wedge J = -8i \, vol_6
\ea
\]
The almost complex structure on the internal manifold is given by
$J_m{}^n=h^{ln}J_{ml}$ ($J^2=-\mathbf{1}$) and can be used to define
holomorphic projectors so that $\Omega$ is a $(3,0)$-form with respect to
$J$. Note, this does not require $J$ to be integrable or the manifold to
be complex.

Some further useful identities are
\be191 \ba{rcl}
\chi^T\gamma_{mnpq}\chi^\star&=&-3J_{[mn}J_{pq]} \ ,\\
\bar \Omega^{mnp}\Omega_{pqr}&=&4\delta^{mn}_{qr}-4J_{[q}{}^{[m}J_{r]}{}^{n]}+
 8i\delta^{[m}_{[q}J_{r]}^{n]}\ , \\
\bar \Omega^{mnp}\Omega_{npq}&=&8\delta^m_q+8iJ_q{}^m \ ,\\
\bar \Omega^{mnp}\Omega_{mnp}&=&48 \ ,\\
J^{mn}J_{[mn}T_{p]}&=&{4\over 3}T_p\ , \\
\bar \Omega^{mnp} \Omega_{[mnp}T_{q]}&=&6(\delta_q^r\!-
\!iJ_q{}^r)T_r\ . 
\ea \ee
The last three can also be
obtained from realizing $\Omega_{ijk}=i\varepsilon_{ijk}$ and
$\varepsilon_{123}=\varepsilon_{\bar 1\bar 2\bar 3}=\sqrt{8}$.

\bigskip


\noindent
{\bf Acknowledgments}

\noindent	
The work of MC was supported in part by the DOE grant
DOE-EY-76-02-3071, NSF grant INT02-03585, and Fay R. and Eugene
L. Langberg endowed Chair. KB would like to thank the theory groups
at the University of Pennsylvania and the Albert-Einstein-Institute
where this work has been initiated.



%

\providecommand{\href}[2]{#2}\begingroup\raggedright\endgroup

\end{document}